\documentclass[aps,prl,preprint,groupedaddress,twocolumn]{revtex4}
\usepackage{amsmath}
\usepackage{amssymb}
\usepackage{graphicx}
\usepackage{epsf}
\usepackage{mathtools}
\begin{document}

% Use the \preprint command to place your local institutional report
% number in the upper righthand corner of the title page in preprint mode.
% Multiple \preprint commands are allowed.
% Use the 'preprintnumbers' class option to override journal defaults
% to display numbers if necessary
%\preprint{}

%Title of paper
\title{Measure of tripartite quantum correlation in multiqubit  symmetric pure state.}

% repeat the \author .. \affiliation  etc. as needed
% \email, \thanks, \homepage, \altaffiliation all apply to the current
% author. Explanatory text should go in the []'s, actual e-mail
% address or url should go in the {}'s for \email and \homepage.
% Please use the appropriate macro foreach each type of information

% \affiliation command applies to all authors since the last
% \affiliation command. The \affiliation command should follow the
% other information
% \affiliation can be followed by \email, \homepage, \thanks as well.
\author{Ram Narayan Deb\\
Email: debramnarayan1@gmail.com}
%\homepage[]{Your web page}
%\thanks{}
%\altaffiliation{}
\affiliation{Chandernagore College, Chandernagore, Hooghly, PIN-712136, West Bengal, India}

%Collaboration name if desired (requires use of superscriptaddress
%option in \documentclass). \noaffiliation is required (may also be
%used with the \author command).
%\collaboration can be followed by \email, \homepage, \thanks as well.
%\collaboration{}
%\noaffiliation

\date{\today}

\begin{abstract}
We propose a direct measure of tripartite quantum correlation in an arbitrary symmetric pure state of N correlated two-level atoms (qubits). We compute the third order moments of the collective pseudo-spin operators in terms of the individual atomic operators of the atoms in the assembly and find that all the bipartite quantum correlation terms among the atoms cancel out leaving only the all possible tripartite quantum correlation terms among them. 
We observe that the third order moments are made up of solely the tripartite quantum correlations among the atoms.
This helps to extract out only the tripartite quantum correlations among the N atoms and propose a measure and quantification of these correlations. We also propose the necessary and sufficient condition for the presence of tripartite quantum correlation in such multi-atomic systems. We conjecture the way of determining the third order moments of the pseudo-spin operators of the two-level atoms in the assembly,
experimentally.\\
{\it keywords: tripartite quantum correlation, two-level atom, pseudo-spin operator} 
\end{abstract}

%\pacs {42.50.-p, 42.50.Dv, 03.65.-w}
% insert suggested keywords - APS authors don't need to do this
%\keywords{}

%\maketitle must follow title, authors, abstract, and keywords
\maketitle

% body of paper here - Use proper section commands
% References should be done using the \cite, \ref, and \label commands
\section{1 INTRODUCTION}
% Put \label in argument of \section for cross-referencing
%\section{\label{}}
Quantum correlation in multi-particle systems is an important topic of research over the last two decades. A lot of work has been done in this direction [1-25]. The works of Wootters 
\cite{Wootters2, Hill}, Peres \cite{Peres} and Horodecki \cite{Horodecki2}, and several other researchers have given a strong understanding of quantum correlation in bipartite states \cite{Horodecki}. But, quantum correlation in multi-particle systems is yet to be completely understood. 

In this paper we consider an arbitrary symmetric pure state  of  $N$ correlated two-level atoms (qubits) and calculate the third order moments of the collective pseudo-spin operators in mutually perpendicular directions in a plane perpendicular to the collective mean pseudo-spin vector. Calculation of third order moments of the collective pseudo-spin operators is necessary because the third order moments of the collective pseudo-spin operators of $N$ two-level atoms reveal the tripartite (three particle) quantum correlations among the atoms. After calculating the third order moments 
we find that all possible bipartite (two particle) correlation terms among the atoms cancel out and we are left with only the tripartite correlation terms. This helps to extract out the tripartite correlation terms among the $N$ atoms and propose a measure of tripartite quantum correlations among the atoms.  The second order moments of the collective pseudo-spin operators are made up of only the bipartite  quantum correlation terms among the atoms. The fourth and higher order moments contain the fourth order and higher order correlation terms among the atoms. So, to concentrate only on the tripartite  correlations among the atoms, we calculate the third order moments of the collective pseudo-spin operators.

The computation of higher order moments and study of higher order squeezing have already been discussed in the context of degenerate parametric down-conversion, harmonic generation and resonance fluoresence from an atom in Ref. \cite{Hong}. In the above mentioned three situations the higher order moments have been computed for the electric field operators of the radiation field. In this paper we calculate third order central moments of the spin operators for the atoms and use it to quantify tripartite quantum correlation. The motivation for the calculation of third order central moments also lies on the fact that, third order central moments solely reveal the tripartite quantum correlations among the atoms.

Here, we would like to say that, in this paper we study the tripartite quantum correlations among the $N$ two-level atoms using the collective spin operators of the atoms as it is the standard technique to study the dynamics of atom-field interactions in terms of these spin operators and the creation and annihilation operators of the radiation field. But, since here we concentrate on the two-level atoms only,
we use the collective spin operators of the atoms only and not the creation and annihilation operators of the radiation field. 

In order to study the quantum correlations among the atoms other physical observables such as the electric dipole moments or electric quadrupole moments of the atoms can also be used. But, since, that approach is quite complicated, we consider the collective spin operators to study the quantum correlations among the atoms in this paper.  

The organization of the paper is as follows. In Sec. 2 we briefly describe a two-level atom, pseudo-spin operators and  moments of the pseudo-spin operators in mutually perpendicular directions in a plane perpendicular to the the mean pseudo-spin vector. In Sec. 3 we present the third order moments of these pseudo-spin operators for three two-level atoms and quantify the amount of tripartite quantum correlation present in such system. In Sec. 4 we extend these ideas in the context of $N$ two-level atoms. In Sec. 5 we conjecture the way of determining the third order moments experimentally.
In Sec. 6 we present the summary and conclusion.  The detail calculations of the third order moments are presented in the Appendix.

\subsection{2 TWO-LEVEL ATOM, PSEUDO-SPIN OPERATORS AND THIRD ORDER CENTRAL MOMENTS}
%\subsubsection{}
An atom has many electronic energy levels, but when it interacts with an external monochromatic electromagnetic field, the atom makes a transition from one of its energy level to the other. In this case, we mainly concentrate on those two energy levels and hence the atom is called as a two-level atom. 
We consider a system of $N$ such two-level atoms.
Now, if among the assembly of $N$ such two-level atoms, the $n$-th atom has the upper and lower energy levels, denoted as 
$|u_n\rangle$ and $|l_n\rangle$, respectively, then, we can construct the pseudo-spin operators (with $\hbar = 1$),
\begin{eqnarray}
\hat{J}_{n_x} &=& (1/2)\big(|u_n\rangle\langle l_n| + |l_n\rangle
\langle u_n|\big),\label{1.1a1}\\
\hat{J}_{n_y} &=& (-i/2)\big(|u_n\rangle\langle l_n| - |l_n\rangle
\langle u_n|\big),\label{1.1a2}\\
\hat{J}_{n_z} &=& 
(1/2)\big(|u_n\rangle\langle u_n| - |l_n\rangle\langle l_n|
\big),
\label{1.1a3}
\end{eqnarray}
such that,
\begin{equation}
 [\hat{J}_{n_x}, \hat{J}_{n_y}] = i\hat{J}_{n_z},
\label{1.2}
\end{equation}
 and two more relations with cyclic changes in $x$, $y$ and $z$ \cite{Itano}. For the entire system of $N$ two-level atoms, we have collective pseudo-spin operators,
\begin{eqnarray}
\hat{J}_x = \sum_{i=1}^{N}\hat{J}_{i_{x}}, 
\hat{J}_y = \sum_{i=1}^{N}\hat{J}_{i_{y}}, 
\hat{J}_z = \sum_{i=1}^{N}\hat{J}_{i_{z}},
\label{1.3}
\end{eqnarray}
where each term in the above summations is multiplied by the identity operators for all of the other atoms \cite{Sakurai}, \cite {Itano}.

The individual atomic operators satisfy
\begin{eqnarray}
\big[\hat{J}_{1_x}, \hat{J}_{2_y}\big] &=& 0,~
\big[\hat{J}_{1_x}, \hat{J}_{1_y}\big] = i\hat{J}_{1_z},\nonumber\\
\big[\hat{J}_{2_x}, \hat{J}_{2_y}\big] &=& i\hat{J}_{2_z},...
\label{1.2a1}
\end{eqnarray}

 As a consequence of these commutation
relations, the collective pseudo-spin operators 
$\hat{J}_x$, $\hat{J}_y$ and $\hat{J}_z$ satisfy,
\begin{equation}
[\hat{J}_x , \hat{J}_y] = i\hat{J}_z
\label{1.2a2}
\end{equation} 
and two more
relations with cyclic changes in $x$, $y$ and $z$.

The simultaneous eigenvectors of $\hat{J}^2 = \hat{J}_x^2
+ \hat{J}_y^2 + \hat{J}_z^2$ and $\hat{J}_z$ are denoted
as $|j,m\rangle$ where
\begin{equation} 
\hat{J}^2|j,m\rangle = j(j+1)|j,m\rangle
\label{1.2a3}
\end{equation}
 and 
\begin{equation}
\hat{J}_z|j,m\rangle = m|j,m\rangle.
\label{1.2a4}
\end{equation} 
The quantum number $j$ is related to the number of atoms 
$N$ as $j = N/2$ and $m = -j, -j+1, ....(j-1), j$.

The use of these collective spin operators and their eigenvectors as discussed in Eqs. (\ref{1.2a2}) to (\ref{1.2a4}) are widely used in the problems of quantum optics [33-36].

The collective quantum state vector for a system of $N$
two-level atoms can be expressed as a linear superposition of $|j,m\rangle$ as  
\begin{equation}
|\psi_j\rangle = \sum_{m=-j}^{j}c_{m}|j,m\rangle.
\label{1.2a5}
\end{equation}
Now, if $X$ is a random variable with its mean value as $\langle X \rangle = m$, then, the quantity $\langle (X - m)^k\rangle$ is called the $k-th$ order central moment of $X$. If $k =2$, then it is the second order central moment of $X$ which is also called the variance of $X$.

In this paper we calculate the third order central moments of the collective pseudo-spin operators in two mutually perpendicular directions in a plane perpendicular to the  collective mean pseudo-spin vector. The motivation of doing so lies on the fact that while investigating spin squeezing in such systems, we calculate the second order central moments 
\begin{equation}
\Delta J_{x,y}^2 = \langle\psi_j|{\hat{J}_{x,y}}^2|\psi_j\rangle - \langle\psi_j|\hat{J}_{x,y}|\psi_j\rangle^2 
\label{2.5a}
\end{equation}     
in a plane perpendicular to the mean spin vector
\begin{equation}
\langle\hat{\mathbf{J}}\rangle = \langle\hat{J}_x\rangle
\hat{i} + \langle\hat{J}_y\rangle\hat{j} + \langle\hat{J}_z\rangle\hat{k},\label{1.4}
\end{equation}
where $\hat{i}$, $\hat{j}$ and $\hat{k}$ are the unit vectors along positive $x$, $y$ and $z$ axes respectively and the expectation values $\langle\hat{J}_x\rangle$, $\langle\hat{J}_y\rangle$ and $\langle\hat{J}_z\rangle$ are with respect to the state in Eq. (\ref{1.2a5}).

. This is because only in that plane the quantum fluctuations in Eqs. (\ref{2.5a}) bring out the original quantum correlations among the atoms \cite{Kitagawa, Wineland2}.
Now, the mean pseudo-spin vector 
$\langle\hat{\mathbf{J}}\rangle$ points in an arbitrary direction in space. Therefore, we conventionally rotate the coordinate system $\{x,y,z\}$ to
$\{x^\prime, y^\prime, z^\prime\}$, such that 
$\langle\hat{\bf{J}}\rangle$ points along the 
$z^\prime$ axis and calculate the second order central moments in 
$\hat{J}_{x^\prime}$ and $\hat{J}_{y^\prime}$ for the state
$|\psi_j\rangle$. These moments are
\begin{equation}
\Delta J_{x^\prime,y^\prime}^2 = \langle\psi_j|{\hat{J}_{x^\prime,y^\prime}}^2|\psi_j\rangle - \langle\psi_j|\hat{J}_{x^\prime,y^\prime}|\psi_j\rangle^2.
\label{2.5}
\end{equation}
Now, in the $x^\prime-y^\prime$ plane
\begin{eqnarray}
\langle\psi_j|\hat{J}_{x^\prime}|\psi_j\rangle = \langle\psi_j|\hat{J}_{y^\prime}|\psi_j\rangle = 0,
\label{2.5a1}
\end{eqnarray}
and hence, these moments reduce to
\begin{equation}
\Delta J_{x^\prime,y^\prime}^2 = \langle\psi_j|{\hat{J}_{x^\prime,y^\prime}}^2|\psi_j\rangle.
\label{2.5a2}
\end{equation}
The third order central moments are defined as
\begin{equation}
\Delta J_{x^\prime,y^\prime}^3 = \langle\psi_j| (\hat{J}_{x^\prime, y^\prime} - \langle \hat{J}_{x^\prime, y^\prime}\rangle)^3|\psi_j\rangle, 
\label{2.5a3}
\end{equation}
which due to Eq. (\ref{2.5a1}) reduces to
\begin{equation}
\Delta J_{x^\prime,y^\prime}^3 = \langle\psi_j|{\hat{J}_{x^\prime,y^\prime}}^3|\psi_j\rangle.
\label{2.5a4}
\end{equation}
A collective state vector $|\alpha\rangle$ for a system of $N$ atoms is said to be quantum mechanically correlated  if 
$|\alpha\rangle$ cannot be expressed as a product
of the individual atomic state vectors, i.e.,
\begin{equation}
|\alpha\rangle \ne |\alpha_1\rangle \otimes|\alpha_2\rangle \otimes....|\alpha_N\rangle,
\label{1.7}
\end{equation}
where $|\alpha_1\rangle$, $|\alpha_2\rangle$...$|\alpha_N\rangle$ are the 
state vectors of the $N$ individual atoms.

We, now, proceed to calculate the third order moments $\Delta J_{x^\prime,y^\prime}^3$ for a system of three two-level atoms and define a parameter to quantify the amount of tripartite quantum correlation in such system.

\subsection{3 TRIPARTITE QUANTUM CORRELATION IN A SYSTEM OF THREE TWO-LEVEL ATOMS}
We consider a system of three two-level atoms.
A symmetric quantum state vector for such a system in the $\{m_1, m_2, m_3\}$ representation can be written as
\begin{eqnarray}
|\psi_3\rangle &=& C_1\bigg\vert\frac{1}{2},\frac{1}{2},\frac{1}{2} \bigg\rangle + \frac{C_2}{\sqrt{3}}\Bigg{[}\bigg\vert-\frac{1}{2},\frac{1}{2},\frac{1}{2} \bigg\rangle\nonumber\\
 &+& \bigg\vert\frac{1}{2},-\frac{1}{2},\frac{1}{2} \bigg\rangle + \bigg\vert\frac{1}{2},\frac{1}{2},-\frac{1}{2} \Bigg\rangle\Bigg{]}
\nonumber\\
&+& \frac{C_3}{\sqrt{3}}\Bigg{[}\bigg\vert-\frac{1}{2},-\frac{1}{2},\frac{1}{2} \bigg\rangle
 + \bigg\vert\frac{1}{2},-\frac{1}{2},-\frac{1}{2} \bigg\rangle\nonumber\\ &+& \bigg\vert-\frac{1}{2},\frac{1}{2},-\frac{1}{2} \Bigg\rangle\Bigg{]} + C_4\bigg\vert-\frac{1}{2},-\frac{1}{2},-\frac{1}{2} \bigg\rangle.\nonumber\\
\label{3.1a}
\end{eqnarray}
 The collective pseudo-spin operators, from Eqs. (\ref{1.3}), are
\begin{equation}
\hat{J}_{x,y,z} = \hat{J}_{1_{x,y,z}} + \hat{J}_{2_{x,y,z}} + \hat{J}_{3_{x,y,z}}.
\label{3.1}
\end{equation}

Now, the mean spin vector $\langle{\bf\hat{J}}\rangle$ points in an arbitrary direction in space. So, assuming that it lies in the first octant of the coordinate system, we perform a rotation of the coordinate system from $\{x, y, z\}$ to $\{x^\prime, y^\prime, z^\prime\}$, such that $\langle\hat{\mathbf{J}}\rangle$ points along the $z^\prime$ axis. The operators
$\{\hat{J}_{x^\prime}, \hat{J}_{y^\prime}, \hat{J}_{z^\prime}\}$ in the rotated frame $\{x^\prime, y^\prime, z^\prime\}$ are related to $\{\hat{J}_x, \hat{J}_y, \hat{J}_z\}$ in the unrotated frame $\{x, y, z\}$ as  

\begin{eqnarray}
\hat{J}_{x^\prime} &=& \hat{J}_x\cos\theta\cos\phi + \hat{J}_y
\cos\theta\sin\phi\nonumber\\ 
&-& \hat{J}_z\sin\theta\label{2.5d1}\\
\hat{J}_{y^\prime} &=& -\hat{J}_x\sin\phi + \hat{J}_y\cos\phi
\label{2.5d2}\\
\hat{J}_{z^\prime} &=& \hat{J}_x\sin\theta\cos\phi + \hat{J}_y
\sin\theta\sin\phi\nonumber\\
&+& \hat{J}_z\cos\theta,\label{2.5d3}
\end{eqnarray}
where,
\begin{eqnarray}
\cos\theta &=& \frac{\langle\hat{J}_z\rangle}
{|\langle\hat{\mathbf{J}}\rangle|}\label{2.5d4}\\
\cos\phi &=& \frac{\langle\hat{J}_x\rangle}{\sqrt{\langle\hat{J}_x\rangle^2 + \langle\hat{J}_y\rangle^2}}.
\label{2.5d5}
\end{eqnarray}
We can observe that for the above choice of $\cos\theta$ 
and $\cos\phi$, we have $\langle\hat{J}_{x^\prime}\rangle = 0$, $\langle\hat{J}_{y^\prime}\rangle = 0$, and the mean spin vector points along the $z^\prime$ axis.

To calculate the third order moment of $\hat{J}_{x^\prime}$ we need to calculate $\hat{J}_{x^\prime}^3$. The calculation of the third order moment $\Delta J_{x^\prime}^3$ has been presented in the Appendix. Here we present the expression of $\Delta J_{x^\prime}^3$ as
\begin{eqnarray}
\Delta J_{x^\prime}^3 &=& 
\frac{1}{|\langle\bf{\hat{J}}\rangle|\big[\langle\hat{J}_x\rangle^2 + \langle\hat{J}_y\rangle^2\big]^{3/2}}\nonumber\\ 
&\times& \sum_{\substack{p,q,r={1}\\ {p\ne q\ne r}}}^{3} \bigg( \langle\hat{J}_{p_x}\hat{J}_{q_x}\hat{J}_{r_x}\rangle\langle\hat{J_z}\rangle^3\langle\hat{J_x}\rangle^3\nonumber\\
&+& \langle\hat{J}_{p_y}\hat{J}_{q_y}\hat{J}_{r_y}\rangle\ \langle\hat{J}_z\rangle^3
\langle\hat{J}_y\rangle^3 - \langle\hat{J}_{p_z}\hat{J}_{q_z}\hat{J}_{r_z}\rangle
\nonumber\\
&\times&\big[\langle\hat{J}_x\rangle^6 + \langle\hat{J}_y\rangle^6 + 3\langle\hat{J}_x\rangle^4\langle\hat{J}_y\rangle^2\nonumber\\  
&+& 
3\langle\hat{J}_x\rangle^2\langle\hat{J}_y\rangle^4  \big]
 - 6 \langle\hat{J}_{p_x}\hat{J}_{q_y}\hat{J}_{r_z}\rangle\nonumber\\
&\times& \big[\langle\hat{J}_x\rangle^2 + \langle\hat{J}_y\rangle^2\big]\langle\hat{J}_z\rangle^2\langle\hat{J}_x\rangle\langle\hat{J}_y\rangle\nonumber\\
 &+& 3\langle
\hat{J}_{p_x}\hat{J}_{q_x}\hat{J}_{r_y}\rangle\langle\hat{J}_z\rangle^3\langle\hat{J}_x\rangle^2\langle\hat{J}_y\rangle\nonumber\\
 &-& 3\langle \hat{J}_{p_x}\hat{J}_{q_x}\hat{J}_{r_z}\rangle
 \big[\langle\hat{J}_x\rangle^2 + \langle\hat{J}_y\rangle^2\big]\langle\hat{J}_x\rangle^2\langle\hat{J}_z\rangle^2\nonumber\\
 &+& 3\langle\hat{J}_{p_x}\hat{J}_{q_y}\hat{J}_{r_y}\rangle\langle\hat{J}_z\rangle^3
\langle\hat{J}_x\rangle\langle\hat{J}_y\rangle^2
\nonumber\\
 &-& 3\langle\hat{J}_{p_y}\hat{J}_{q_y}\hat{J}_{r_z}\rangle\big[\langle\hat{J}_x\rangle^2 + \langle\hat{J}_y\rangle^2\big]\langle\hat{J}_y\rangle^2\langle\hat{J}_z\rangle^2\nonumber\\
 &+& 3 \langle\hat{J}_{p_x}\hat{J}_{q_z}\hat{J}_{r_z}\rangle\langle\hat{J}_x\rangle\langle\hat{J}_z\rangle\big[\langle\hat{J}_x\rangle^4 + \langle\hat{J}_y\rangle^4\nonumber\\ 
&+& 2\langle\hat{J}_x\rangle^2\langle\hat{J}_y\rangle^2\big]
 + 3 \langle\hat{J}_{p_y}\hat{J}_{q_z}\hat{J}_{r_z}\rangle\langle\hat{J}_y\rangle\langle\hat{J}_z\rangle\nonumber\\
&\times& \big[\langle\hat{J}_x\rangle^4 + \langle\hat{J}_y\rangle^4 + 2\langle\hat{J}_x\rangle^2\langle\hat{J}_y\rangle^2\big] \bigg).\nonumber\\
\label{3.30a3b5}
\end{eqnarray}
The terms $\langle\hat{J}_{p_x}\hat{J}_{q_x}\hat{J}_{r_x}\rangle$, $\langle\hat{J}_{p_y}\hat{J}_{q_y}\hat{J}_{r_y}\rangle$,......$\langle\hat{J}_{p_y}\hat{J}_{q_z}\hat{J}_{r_z}\rangle$ in the above equation are the three-particle correlation terms.
So, we observe that $\Delta J_{x^\prime}^3$ are made up of the three-particle correlation terms and all the two-particle correlation terms get cancelled out. 

Similarly the expression of the third order moment in $J_{y^\prime}$ is 
 
\begin{eqnarray}
\Delta J_{y^\prime}^3 &=& 
\frac{1}{\big[\langle\hat{J}_x\rangle^2 + \langle\hat{J}_y\rangle^2\big]^{3/2}} 
\sum_{\substack{p,q,r={1}\\ {p\ne q\ne r}}}^{3}\nonumber\\
&&\bigg( -\langle\hat{J}_{p_x}\hat{J}_{q_x}\hat{J}_{r_x}\rangle\langle\hat{J_y}\rangle^3 + \langle\hat{J}_{p_y}\hat{J}_{q_y}\hat{J}_{r_y}\rangle \nonumber\\
&\times&\langle\hat{J}_x\rangle^3
+ 3\langle
\hat{J}_{p_x}\hat{J}_{q_x}\hat{J}_{r_y}\rangle\langle\hat{J}_x\rangle\langle\hat{J}_y\rangle^2 \nonumber\\
&-& 3\langle\hat{J}_{p_x}\hat{J}_{q_y}\hat{J}_{r_y}\rangle\langle\hat{J}_x\rangle^2\langle\hat{J}_y\rangle\bigg) 
\label{3.31b5}
\end{eqnarray}
While calculating $\Delta J_{y^\prime}^3$, we find that all the bipartite correlation terms cancel out and it is solely made up of tripartite correlation terms like the previous case.

Since the quantum state vector in Eq. (\ref{3.1a}) is a symmetric state
 where all the atoms have been treated on equal footing, then, we have
\begin{eqnarray}
\langle\hat{J}_{1_x}\rangle &=& \langle\hat{J}_{2_x}\rangle = \langle\hat{J}_{3_x}\rangle\label{3.33}\\
\langle\hat{J}_{1_y}\rangle &=& \langle\hat{J}_{2_y}\rangle = \langle\hat{J}_{3_y}\rangle\label{3.34}\\
\langle\hat{J}_{1_z}\rangle &=& \langle\hat{J}_{2_z}\rangle = \langle\hat{J}_{3_z}\rangle.
\label{3.35}
\end{eqnarray}
In this case, using Eqs. (\ref{3.33}), (\ref{3.34}) and, (\ref{3.35}) in Eq. (\ref{3.30a3b5}) and (\ref{3.31b5}), the third order moments of $\hat{J}_{x^\prime}$ and $\hat{J}_{y^\prime}$, reduce to

\begin{eqnarray}
\Delta J_{x^\prime}^3 &=& 
\frac{1}{|\langle\bf{\hat{J}_1}\rangle|\big[\langle\hat{J}_
{1_x}\rangle^2 + \langle\hat{J}_{1_y}\rangle^2\big]^{3/2}}\nonumber\\ 
&\times& \sum_{\substack{p,q,r={1}\\ {p\ne q\ne r}}}^{3} \bigg( \langle\hat{J}_{p_x}\hat{J}_{q_x}\hat{J}_{r_x}\rangle\langle\hat{J_{1_z}}\rangle^3
\langle\hat{J_{1_x}}\rangle^3\nonumber\\
&+& \langle\hat{J}_{p_y}\hat{J}_{q_y}\hat{J}_{r_y}\rangle\ \langle\hat{J}_
{1_z}\rangle^3
\langle\hat{J}_{1_y}\rangle^3 - \langle\hat{J}_{p_z}\hat{J}_{q_z}\hat{J}_{r_z}\rangle\nonumber\\
&\times&\big[\langle\hat{J}_{1_x}\rangle^6 + \langle\hat{J}_{1_y}\rangle^6 
+ 3\langle\hat{J}_{1_x}\rangle^4\langle\hat{J}_{1_y}\rangle^2\nonumber\\  
&+& 3\langle\hat{J}_{1_x}\rangle^2\langle\hat{J}_{1_y}\rangle^4  \big]
- 6 \langle\hat{J}_{p_x}\hat{J}_{q_y}\hat{J}_{r_z}\rangle\nonumber\\ 
&\times& \big[\langle\hat{J}_{1_x}\rangle^2 + \langle\hat{J}_{1_y}\rangle^2\big]\langle\hat{J}_{1_z}\rangle^2\langle\hat{J}
_{1_x}\rangle\langle\hat{J}_{1_y}\rangle\nonumber\\
&+& 3\langle
\hat{J}_{p_x}\hat{J}_{q_x}\hat{J}_{r_y}\rangle\langle\hat{J}_
{1_z}\rangle^3\langle\hat{J}_{1_x}\rangle^2\langle\hat{J}_
{1_y}\rangle\nonumber\\
 &-& 3\langle \hat{J}_{p_x}\hat{J}_{q_x}\hat{J}_{r_z}\rangle
 \big[\langle\hat{J}_{1_x}\rangle^2 + \langle\hat{J}_
{1_y}\rangle^2\big]\langle\hat{J}_{1_x}\rangle^2
\nonumber\\
&\times& \langle\hat{J}_
{1_z}\rangle^2 + 3\langle\hat{J}_{p_x}\hat{J}_{q_y}\hat{J}_{r_y}\rangle\nonumber\\
&\times&\langle\hat{J}_{1_z}\rangle^3\langle\hat{J}_{1_x}\rangle\langle\hat{J}_{1_y}\rangle^2 - 3\langle\hat{J}_{p_y}\hat{J}_{q_y}\hat{J}_{r_z}
\rangle\nonumber\\
&\times& \big[\langle\hat{J}_{1_x}\rangle^2 + \langle\hat{J}_{1_y}\rangle^2\big]
\langle\hat{J}_{1_y}\rangle^2\langle\hat{J}_{1_z}\rangle^2\nonumber\\
 &+& 3 \langle\hat{J}_{p_x}\hat{J}_{q_z}\hat{J}_{r_z}\rangle\langle\hat{J}_{1_x}\rangle\langle\hat{J}_{1_z}\rangle\big[\langle\hat{J}_{1_x}\rangle^4
+ \langle\hat{J}_{1_y}\rangle^4\nonumber\\
&+& 2\langle\hat{J}_{1_x}\rangle^2\langle\hat{J}_{1_y}\rangle^2\big] + 3 \langle\hat{J}_{p_y}\hat{J}_{q_z}\hat{J}_{r_z}\rangle\langle\hat{J}_{1_y}\rangle\langle\hat{J}_{1_z}
\rangle\nonumber\\
&\times&\big[\langle\hat{J}_{1_x}\rangle^4 + \langle\hat{J}_
{1_y}\rangle^4 + 2\langle\hat{J}_{1_x}\rangle^2\langle\hat{J}_
{1_y}\rangle^2\big] \bigg)\nonumber\\
\label{3.30a4}
\end{eqnarray}
and
\begin{eqnarray}
\Delta J_{y^\prime}^3 &=& 
\frac{1}{\big[\langle\hat{J}_{1_x}\rangle^2 + \langle\hat{J}_
{1_y}\rangle^2\big]^{3/2}} 
\sum_{\substack{p,q,r={1}\\ {p\ne q\ne r}}}^{3}\nonumber\\
&&\bigg( -\langle\hat{J}_{p_x}\hat{J}_{q_x}\hat{J}_{r_x}\rangle\langle\hat{J_{1_y}}\rangle^3 + \langle\hat{J}_{p_y}\hat{J}_{q_y}\hat{J}_{r_y}\rangle \nonumber\\
&\times&\langle\hat{J}_{1_x}\rangle^3
+ 3\langle
\hat{J}_{p_x}\hat{J}_{q_x}\hat{J}_{r_y}\rangle\langle\hat{J}_
{1_x}\rangle\langle\hat{J}_{1_y}\rangle^2 \nonumber\\
&-& 3\langle\hat{J}_{p_x}\hat{J}_{q_y}\hat{J}_{r_y}\rangle\langle\hat{J}_{1_x}\rangle^2\langle\hat{J}_{1_y}\rangle\bigg). 
\label{3.36}
\end{eqnarray}

Now, if the three atoms are uncorrelated, then
\begin{eqnarray}
\langle\hat{J}_{p_x}\hat{J}_{q_x}\hat{J}_{r_x}\rangle &=& \langle\hat{J}_{p_x}\rangle\langle\hat{J}_{q_x}\rangle
\langle\hat{J}_{r_x}\rangle\label{3.37}\\
\langle\hat{J}_{p_y}\hat{J}_{q_y}\hat{J}_{r_y}\rangle &=& \langle\hat{J}_{p_y}\rangle\langle\hat{J}_{q_y}\rangle\langle\hat{J}_{r_y}\rangle\label{3.38}\\
\langle\hat{J}_{p_z}\hat{J}_{q_z}\hat{J}_{r_z}\rangle &=& \langle\hat{J}_{p_z}\rangle\langle\hat{J}_{q_z}\rangle\langle\hat{J}_{r_z}\rangle\label{3.39}\\
\langle\hat{J}_{p_x}\hat{J}_{q_y}\hat{J}_{r_z}\rangle &=& \langle\hat{J}_{p_x}\rangle\langle\hat{J}_{q_y}\rangle\langle\hat{J}_{r_z}\rangle\label{3.40}\\ 
\langle\hat{J}_{p_x}\hat{J}_{q_x}\hat{J}_{r_y}\rangle &=& \langle\hat{J}_{p_x}\rangle\langle\hat{J}_{q_x}\rangle\langle\hat{J}_{r_y}\rangle\label{3.41}\\
\langle \hat{J}_{p_x}\hat{J}_{q_x}\hat{J}_{r_z}\rangle &=& \langle \hat{J}_{p_x}\rangle\langle\hat{J}_{q_x}\rangle\langle\hat{J}_{r_z}\rangle\label{3.42}\\
\langle\hat{J}_{p_x}\hat{J}_{q_y}\hat{J}_{r_y}\rangle &=& \langle\hat{J}_{p_x}\rangle\langle\hat{J}_{q_y}\rangle\langle\hat{J}_{r_y}\rangle\label{3.43}\\
\langle\hat{J}_{p_y}\hat{J}_{q_y}\hat{J}_{r_z}\rangle &=& \langle\hat{J}_{p_y}\rangle\langle\hat{J}_{q_y}\rangle\langle\hat{J}_{r_z}\rangle\label{3.44}\\
\langle\hat{J}_{p_x}\hat{J}_{q_z}\hat{J}_{r_z}\rangle &=&\langle\hat{J}_{p_x}\rangle\langle\hat{J}_{q_z}\rangle\langle\hat{J}_{r_z}\rangle\label{3.45}\\
\langle\hat{J}_{p_y}\hat{J}_{q_z}\hat{J}_{r_z}\rangle &=& \langle\hat{J}_{p_y}\rangle\langle\hat{J}_{q_z}\rangle\langle\hat{J}_{r_z}\rangle
\label{3.46}
\end{eqnarray}
Using Eqs. (\ref{3.37}-\ref{3.46}) and 
Eqs. (\ref{3.33}-\ref{3.35}) in Eqs. (\ref{3.30a4}) and (\ref{3.36}), we observe that all the correlation terms on the right hand side of Eqs. (\ref{3.30a4}) and (\ref{3.36}) vanish, and we get
\begin{eqnarray}
\Delta J_{x^\prime}^3 &=& 0\\
\Delta J_{y^\prime}^3 &=& 0.
\end{eqnarray}

 This confirms that the third order moments of $\hat{J}_{x^\prime}$ and $\hat{J}_{y^\prime}$ are the direct measures of tripartite correlations among the atoms.

So, now, we construct a parameter that quantifies the amount of tripartite correlations in this system of three two-level atoms.
Since $\Delta J_{x^\prime}^3$ and $\Delta J_{y^\prime}^3$ may be both positive or negative, and, to treat the tripartite correlations in both $x^\prime$ and $y^\prime$ quadratures on equal footing, we define the tripartite quantum correlation parameter as the  mean squared value of  
$\Delta J_{x^\prime}^3$ and $\Delta J_{y^\prime}^3$ as
\begin{equation}
S = \frac{1}{2}\Big[\big(\Delta J_{x^\prime}^3\big)^2 + \big(\Delta J_{y^\prime}^3\big)^2\Big].
\label{3.47}
\end{equation}
Since, $\Delta J_{x^\prime}^3$ and $\Delta J_{y^\prime}^3$ are solely made up of tripartite quantum correlations only, we may take the numerical value of $S$ as a measure of the amount of tripartite quantum correlation present in the system. Whenever 
\begin{equation}
S = 0,
\label{3.48}
\end{equation}
there is no tripartite quantum correlation in the system.

\begin{figure}
\begin{center}
\includegraphics[width=6cm]{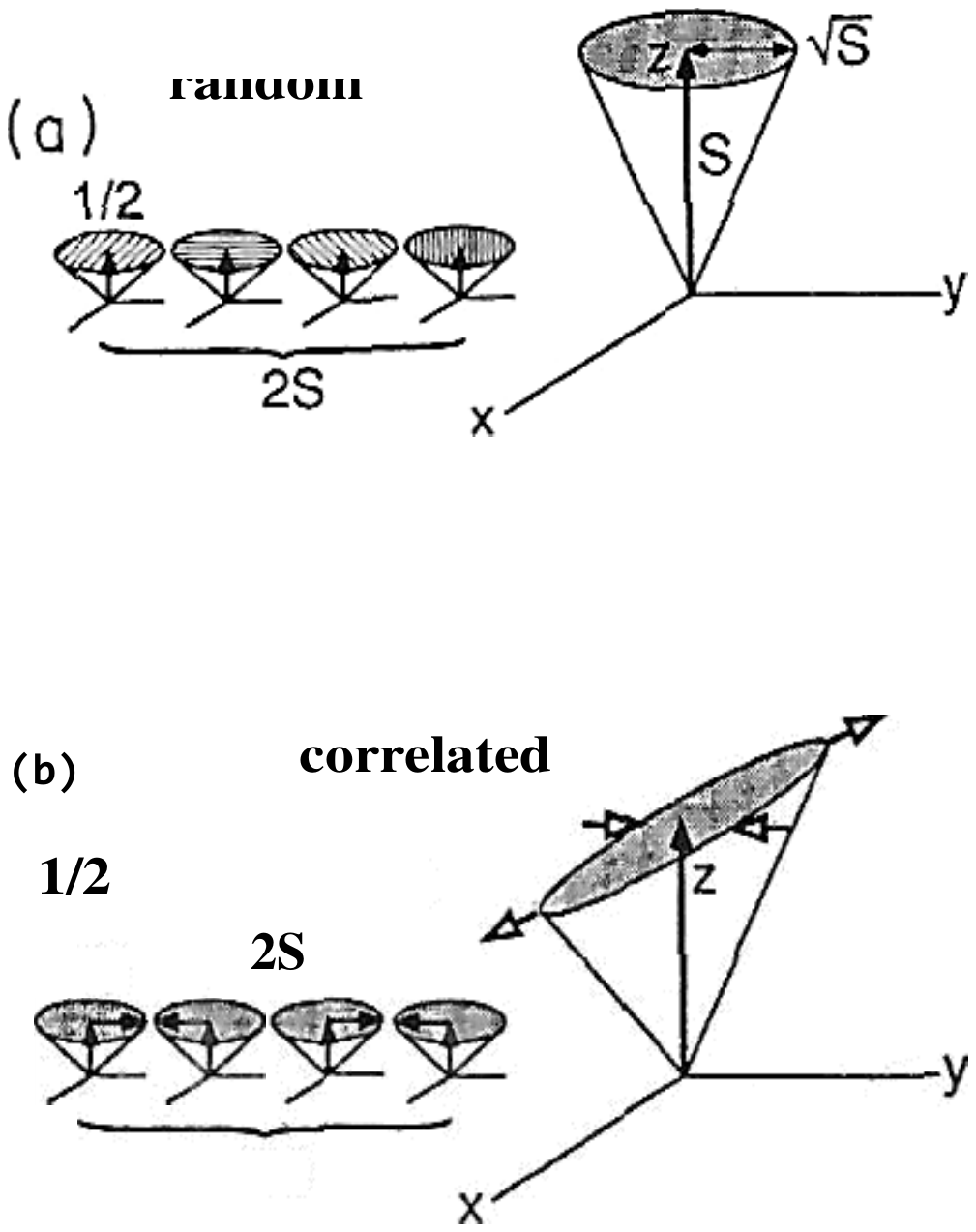}
\caption{Schematic illustrations of S-spin states in terms of 2S = N individual 1/2 spins. (a) Symmetric distribution of  $J_{x^\prime}$ and $J_{y^\prime}$ about the collective mean pseudo-spin vector of 2S = N uncorrelated 1/2 spins. (b) Asymmetric distribution of $J_{x^\prime}$ and $J_{y^\prime}$ about the collective mean pseudo-spin vector of 2S = N correlated 1/2 spins.}
\end{center}
\label {fig1}
\end{figure}

The necessary and sufficient condition for the presence of tripartite quantum correlation in the system is that,
\begin{equation}
S > 0.
\label{3.48}
\end{equation}
We prove it in this way. Whenever there is tripartite quantum correlation in the system, the conditions in Eqs. (\ref{3.37}) to (\ref{3.46}) are not satisfied. This makes $\Delta J_{x^\prime}^3$ and $\Delta J_{y^\prime}^3$ non-zero, implying that
 $S > 0$. So, $S > 0$ is the necessary condition for the presence of tripartite quantum correlation in the system. The condition $S > 0$ is sufficient also for the presence of tripartite quantum correlation in the system, because, if $S > 0$, either $\Delta J_{x^\prime}^3$ or $\Delta J_{y^\prime}^3$ or both of them are non-zero. In this case all the conditions in Eqs. (\ref{3.37}) to (\ref{3.46}) are not satisfied and hence, there is tripartite quantum correlation in the system.   
 
This can be understood physically in the following way. When the $N$ two-level atoms are uncorrelated, the collective pseudo-spin operators $J_{x^\prime}$ and 
$J_{y^\prime}$ have symmetric distribution about the mean pseudo-spin vector, which lies along the $z^\prime$ axis. This has been shown in Fig. 1(a). But, when the $N$ two level atoms are correlated the $J_{x^\prime}$ and $J_{y^\prime}$ have asymmetric distribution about the mean pseudo-spin vector. This has been shown in Fig. 1(b). These figures have been taken from the paper by Kitagawa and Ueda \cite{Kitagawa}. Now, the third order moments of any distribution function measures the departure from symmetry of the distribution. So, when the atoms are correlated having asymmetric distribution of $J_{x^\prime}$ and $J_{y^\prime}$,  the third order moments are non-zero, thus, making 
$S > 0$.

Now, we present some numerical values of $S$ for various choices of the constants $C_1$, $C_2$, $C_3$ and $C_4$ in the quantum state represented in Eq. (\ref{3.1a}). The various choices of these constants produces different quantum states of the three two-level atoms.
\begin{table}[ht]
\caption{Values of $C_1$, $C_2$, $C_3$, $C_4$ and $S$ for 
  the number of atoms $N = 3$ (with $\hbar = 1$).}
\centering
\begin{tabular}{c c c c c}
\hline\hline
$C_1$ & $C_2$ & $C_3$ & $C_4$ & $S$ \\ [0.5ex]
 1 & 0 & 0 & 0 & 0\\    
 0 & 0 & 0 & 1 & 0\\    
 0.316 & 0.316  & 0.447 & 0.775 &    0.0035\\    
 0.447 & 0.548  &     0.447 &      0.548 & 0.0024\\    
 [1 ex]   
\hline
\end{tabular}
\label{table:quant}
\end{table}
We observe that, when $C_1 = 1$ and all other constants are zero, the value of $S$ is zero. This is acceptable because when $C_1 = 1$ and all other constants are zero, all the atoms are in their upper states which represent an uncorrelated state. Similarly, when $C_4 = 1$ and all other constants are zero, all the atoms are in their lower states which also is an uncorrelated state. Therefore, for the above two cases $S = 0$. The numerical values in the table are rounded off values.  

We now proceed to extend these ideas for a system of $N$ two-level atoms.

\subsection{4 TRIPARTITE QUANTUM CORRELATION IN A SYSTEM OF N TWO-LEVEL ATOMS}

We consider $N$ two-level atoms in an arbitrary symmetric pure state. The quantum state vector for this system in $m_1, m_2, m_3,.....m_N$ representation is 

\begin{eqnarray}
|\Psi_N\rangle &=& G_1\bigg\vert\frac{1}{2},\frac{1}{2},....\frac{1}{2}
\bigg\rangle
+ \frac{G_2}{\sqrt{{}^NC_1}}\Bigg[\bigg\vert-\frac{1}{2},\frac{1}{2},\nonumber\\
&&\frac{1}{2},....\frac{1}{2}
\bigg\rangle + \bigg\vert\frac{1}{2},-\frac{1}{2},\frac{1}{2},....\frac{1}{2}
\bigg\rangle\nonumber\\
&+& ....\bigg\vert\frac{1}{2},\frac{1}{2},\frac{1}{2},....
-\frac{1}{2}\bigg\rangle\Bigg]\nonumber\\ 
&+& \frac{G_3}{\sqrt{{}^NC_2}}\Bigg[\bigg\vert-\frac{1}{2},-\frac{1}{2},\frac{1}{2},....\frac{1}{2}
\bigg\rangle \nonumber\\
&+& \bigg\vert-\frac{1}{2},\frac{1}{2},-\frac{1}{2},....\frac{1}{2}
\bigg\rangle + .... \bigg\vert\frac{1}{2},\frac{1}{2}
\nonumber\\
&&,....
-\frac{1}{2},-\frac{1}{2}\bigg\rangle\Bigg]
+ ............\nonumber\\
&&G_{N+1}\bigg\vert-\frac{1}{2},-\frac{1}{2},-
\frac{1}{2},....-\frac{1}{2}\bigg\rangle,
\label{4.1}
\end{eqnarray}
where $G_1$, $G_2$, ..., $G_{N+1}$ are constants and
 ${}^N C_r$ is given as
\begin{equation}
{}^NC_r = \frac{N!}{r!(N-r)!}.
\label{4.2}
\end{equation}

The third order moments in $\hat{J}_{x^\prime}$ and 
$\hat{J}_{y^\prime}$ for this state can be written in analogy to Eqs. (\ref{3.30a4}) and (\ref{3.36}) as
\begin{eqnarray}
\Delta J_{x^\prime}^3 &=& 
\frac{1}{|\langle\bf{\hat{J}_1}\rangle|\big[\langle\hat{J}_
{1_x}\rangle^2 + \langle\hat{J}_{1_y}\rangle^2\big]^{3/2}}\nonumber\\ 
&\times& \sum_{\substack{p,q,r={1}\\ {p\ne q\ne r}}}^{N} \bigg( \langle\hat{J}_{p_x}\hat{J}_{q_x}\hat{J}_{r_x}\rangle\langle\hat{J_{1_z}}\rangle^3
\langle\hat{J_{1_x}}\rangle^3\nonumber\\
&+& \langle\hat{J}_{p_y}\hat{J}_{q_y}\hat{J}_{r_y}\rangle\ \langle\hat{J}_
{1_z}\rangle^3
\langle\hat{J}_{1_y}\rangle^3 - \langle\hat{J}_{p_z}\hat{J}_{q_z}\hat{J}_{r_z}\rangle\nonumber\\
&\times&\big[\langle\hat{J}_{1_x}\rangle^6 + \langle\hat{J}_{1_y}\rangle^6 
+ 3\langle\hat{J}_{1_x}\rangle^4\langle\hat{J}_{1_y}\rangle^2\nonumber\\  
&+& 3\langle\hat{J}_{1_x}\rangle^2\langle\hat{J}_{1_y}\rangle^4  \big]
- 6 \langle\hat{J}_{p_x}\hat{J}_{q_y}\hat{J}_{r_z}\rangle\nonumber\\ 
&\times& \big[\langle\hat{J}_{1_x}\rangle^2 + \langle\hat{J}_{1_y}\rangle^2\big]\langle\hat{J}_{1_z}\rangle^2\langle\hat{J}
_{1_x}\rangle\langle\hat{J}_{1_y}\rangle\nonumber\\
&+& 3\langle
\hat{J}_{p_x}\hat{J}_{q_x}\hat{J}_{r_y}\rangle\langle\hat{J}_
{1_z}\rangle^3\langle\hat{J}_{1_x}\rangle^2\langle\hat{J}_
{1_y}\rangle\nonumber\\
 &-& 3\langle \hat{J}_{p_x}\hat{J}_{q_x}\hat{J}_{r_z}\rangle
 \big[\langle\hat{J}_{1_x}\rangle^2 + \langle\hat{J}_
{1_y}\rangle^2\big]\langle\hat{J}_{1_x}\rangle^2
\nonumber\\
&\times& \langle\hat{J}_
{1_z}\rangle^2 + 3\langle\hat{J}_{p_x}\hat{J}_{q_y}\hat{J}_{r_y}\rangle\nonumber\\
&\times&\langle\hat{J}_{1_z}\rangle^3\langle\hat{J}_{1_x}\rangle\langle\hat{J}_{1_y}\rangle^2 - 3\langle\hat{J}_{p_y}\hat{J}_{q_y}\hat{J}_{r_z}
\rangle\nonumber\\
&\times& \big[\langle\hat{J}_{1_x}\rangle^2 + \langle\hat{J}_{1_y}\rangle^2\big]
\langle\hat{J}_{1_y}\rangle^2\langle\hat{J}_{1_z}\rangle^2\nonumber\\
 &+& 3 \langle\hat{J}_{p_x}\hat{J}_{q_z}\hat{J}_{r_z}\rangle\langle\hat{J}_{1_x}\rangle\langle\hat{J}_{1_z}\rangle\big[\langle\hat{J}_{1_x}\rangle^4
+ \langle\hat{J}_{1_y}\rangle^4\nonumber\\
&+& 2\langle\hat{J}_{1_x}\rangle^2\langle\hat{J}_{1_y}\rangle^2\big] + 3 \langle\hat{J}_{p_y}\hat{J}_{q_z}\hat{J}_{r_z}\rangle\langle\hat{J}_{1_y}\rangle\langle\hat{J}_{1_z}
\rangle\nonumber\\
&\times&\big[\langle\hat{J}_{1_x}\rangle^4 + \langle\hat{J}_
{1_y}\rangle^4 + 2\langle\hat{J}_{1_x}\rangle^2\langle\hat{J}_
{1_y}\rangle^2\big] \bigg).\nonumber\\
\label{3.30b4}
\end{eqnarray}
and,
\begin{eqnarray}
\Delta J_{y^\prime}^3 &=& 
\frac{1}{\big[\langle\hat{J}_{1_x}\rangle^2 + \langle\hat{J}_
{1_y}\rangle^2\big]^{3/2}} 
\sum_{\substack{p,q,r={1}\\ {p\ne q\ne r}}}^{N}\nonumber\\
&&\bigg( -\langle\hat{J}_{p_x}\hat{J}_{q_x}\hat{J}_{r_x}\rangle\langle\hat{J_{1_y}}\rangle^3 + \langle\hat{J}_{p_y}\hat{J}_{q_y}\hat{J}_{r_y}\rangle \nonumber\\
&\times&\langle\hat{J}_{1_x}\rangle^3
+ 3\langle
\hat{J}_{p_x}\hat{J}_{q_x}\hat{J}_{r_y}\rangle\langle\hat{J}_
{1_x}\rangle\langle\hat{J}_{1_y}\rangle^2 \nonumber\\
&-& 3\langle\hat{J}_{p_x}\hat{J}_{q_y}\hat{J}_{r_y}\rangle\langle\hat{J}_{1_x}\rangle^2\langle\hat{J}_{1_y}\rangle\bigg). 
\label{3.36b5}
\end{eqnarray}

respectively, where the upper index $3$ in the summations in Eqs (\ref{3.30a4}) and (\ref{3.36}) has been replaced by $N$.

Now, if all the $N$ atoms are uncorrelated, then, the conditions of Eqs. (\ref{3.37}) to (\ref{3.46}) are satisfied, and, we get
\begin{equation}
\Delta J_{x^\prime}^3 = \Delta J_{y^\prime}^3 = 0.
\label{4.8}
\end{equation}

Thus, for the system of $N$ two-level atoms (qubits), we observe that the third order moments of $\hat{J}_{x^\prime}$ and $\hat{J}_{y^\prime}$ are directly related to the all possible tripartite correlations among the atoms.

In this case also, we define the measure of tripartite quantum correlation among the atoms as the mean squared value of  
$\Delta J_{x^\prime}^3$ and  $\Delta J_{y^\prime}^3$ as
\begin{equation}
S = \frac{1}{2}\Big[\big(\Delta J_{x^\prime}^3\big)^2 + \big(\Delta J_{y^\prime}^3\big)^2\Big].
\label{4.9}
\end{equation}
Like our previous case the necessary and sufficient condition for the presence of tripartite quantum correlation in this system of $N$ two-level atoms is
\begin{equation}
S > 0.
\label{4.10}
\end{equation} 

The tripartite quantum correlation parameter $S$ can be calculated analytically for several systems of two-level atoms such as Atomic 
Schr$\ddot{o}$dinger Cat states, $N$ two-level atoms interacting with the squeezed vacuum state of the radiation field, two-level atoms in optical cavities etc. But, the analytical calculations of $S$ for only a system of four two-level atoms is extremely cumbersome and tedious. The analytical expression of $S$ is also too large to be incorporated in this paper. It is wise to calculate $S$ numerically. 

Here, we would like to mention that, if we calculate the higher order central moments of the collective pseudo-spin operators, we obtain the higher order quantum correlation terms among the atoms. For example, if we calculate the $k-th$ order central moments of these spin operators we get the $k-th$ order quantum correlation among the atoms. Now, the exact mathematical relationship between the $k-th$ order moments and $k-th$ order correlations can be obtained 
only after we calculate and observe the exact mathematical forms of these moments.

Our approach of measuring tripartite quantum correlations in symmetric multiqubit pure states can be extended to investigate quantum correlation in symmetric multiqubit mixed states where the third order central moments of the spin operators are to be calculated using density operators.

\section{5 CONJECTURE FOR DETERMINING THIRD ORDER MOMENTS BY EXPERIMENTAL TECHNIQUE}

Computation of higher order moments and study of higher order squeezing has been described in Ref. \cite{Hong} in the context of 
degenerate parametric down-conversion, harmonic generation and resonance fluoresence from an atom. In the above mentioned three situations the higher order moments have been computed for the electric field operators of the radiation field.
 Below we describe the area of experiment where the third order moments of the angular momentum or pseudo-spin operators of two-level atoms can be determined.

In Ref. \cite{Wineland2}, Wineland et. al described population spectroscopy of $N$ two-level atoms. They localize an ensemble of $N$ identical two-level atoms in a trap. They denote the upper and lower energy levels of the atom as $|+\frac{1}{2}\rangle$ and $|-\frac{1}{2}\rangle$ respectively. They apply a classical radiation called the clock radiation of frequency 
$\omega$ to the atoms. As a result an atom acquires a coherent superposition state $C_1 |+\frac{1}{2}\rangle + C_2 |-
\frac{1}{2}\rangle$. Then they detect the number of atoms in 
$|+\frac{1}{2}\rangle$ state, which they denote as $N_{+}$. In the detection process each atom is projected either in state 
$|+\frac{1}{2}\rangle$ or $|-\frac{1}{2}\rangle$. So, there occurs a projection noise of the atoms. Due to the analogy between an individual two-level system interacting with radiation with that of the dynamics of a spin-half particle in a magnetic field \cite{Feynman}, the quantum projection noise is proportional to 
$\Delta J_z = \sqrt{\langle\hat{J}_{z}^2\rangle - \langle\hat{J}_z\rangle^2}$, where $\hat{J}_z$ is the collective angular momentum operator of $N$ atoms, which we call as the pseudo-spin operator in Eq. (\ref{1.3}) in our paper. By making measurements of the population in $|+\frac{1}{2}\rangle$
state $M$ times the average $(N_+)_{M}$ is calculated for various values of $\omega$, Consequently, a resonance curve is obtained as a function of $\omega$. For a particular value of 
$\omega$, the deviation of the apparent position of the curve  from the true curve $\langle N_+\rangle$ is given by
\begin{equation}
\delta\omega_M = [{(N_{+})}_{M} - \langle N_+\rangle]/(\partial\langle N_+\rangle/\partial\omega).
\end{equation}    
The magnitude of the rms fluctuation of $\delta\omega$ for repeated measurements of $N_+$ at a particular value of $\omega$ is given by
\begin{eqnarray}
|\Delta\omega| &=& \Delta N_{+} (t_f)/ |\partial\langle N_{+}\rangle/\partial\omega|\nonumber\\
&=& \Delta {J_z}{(t_f)}/ |\partial\langle {J_z}{(t_f)}
\partial\omega|,
\end{eqnarray}
where $t_f$ is the "final" time corresponding to the time just after the clock radiation is applied. Then they describe population spectroscopy using Ramsey method. In Sec. V of that paper, where they describe spectroscopy of correlated particles, they discuss spin squeezing, thereby calculating the variance $\Delta J_{\perp}$
in a plane perpendicular to the mean spin vector 
$\langle\hat{\mathbf{J}}\rangle$. So they have determined the second order moments of the angular momentum or pseudo-spin operators $\hat{J}_x$, $\hat{J}_y$, and $\hat{J}_z$ experimentally.  Now, to calculate higher order squeezing one needs to calculate higher order moments. Also to calculate the skewness of the resonance curve of Sec. V of that paper one needs to calculate the third order moments of the angular momentum or pseudo-spin operators. Here we conjecture that with the same experimental technique as described in the paper of Wineland et. al \cite{Wineland2} the third order moments in the angular momentum or pseudo-spin operators can be determined, and as proposed by us we can calculate the amount of tripartite quantum correlation present in the system.

\subsection{6 SUMMARY AND CONCLUSION}

We proposed a measure of tripartite quantum correlation present in an arbitrary symmetric pure state of $N$ two-level atoms. We calculated the third order moments of the collective pseudo-spin operators in mutually orthogonal directions in a plane perpendicular to the mean pseudo-spin vector operator
$\langle\bf\hat{J}\rangle$. We calculated the third order moments in terms of the individual atomic operators. We found that in the expressions of third order moments all the bipartite correlation terms cancel out. These moments depend solely on the tripartite quantum correlation terms among the atoms. This helped to construct a measure of tripartite quantum correlations present among the atoms. Since, the two third order moments $\Delta J_{x^\prime}^3$ and 
$\Delta J_{y^\prime}^3$ may be positive or negative, and, also to treat the two moments on equal footing, we defined the measure of tripartite quantum correlation as the mean squared value of the two moments. We also established the necessary and sufficient condition for the presence of tripartite quantum correlation in such multi-atomic systems. We hope that our study may provide a new insight in the tripartite quantum correlations among $N$ two-level atoms in an arbitrary symmetric pure state. We conjecture the experiment where the third order moments of the angular momentum or pseudo-spin operators can be determined.

\subsection{7 APPENDIX}

Here we show the calculations of the third order moments $\Delta {J}_{x^\prime}^3$ and $\Delta{J}_{y^\prime}^3$ for three two-level atoms.

Using Eq. (\ref{2.5d1}), we obtain
\begin{eqnarray}
{\hat{J}_{x^\prime}}^3 &=& \hat{J}_{x}^3\cos^3\theta\cos^3\phi + \hat{J}_{x}^2\hat{J}_{y}\cos^3\theta\sin\phi\nonumber\\
&\times& \cos^2\phi
-\hat{J}_{x}^2\hat{J}_{z}\sin\theta\cos^2\theta\cos^2\phi\nonumber\\
 &+& \hat{J}_{y}^2\hat{J}_{x}\cos^3\theta\sin^2\phi\cos\phi
+ \hat{J}_{y}^3\cos^3\theta\nonumber\\
&\times& \sin^3\phi 
- \hat{J}_{y}^2\hat{J}_{z}\sin\theta\cos^2\theta\sin^2\phi\nonumber\\
&+& \hat{J}_{z}^2\hat{J}_{x}\sin^2\theta\cos\theta\cos\phi\nonumber\\
&+& \hat{J}_{z}^2\hat{J}_{y}\sin^2\theta\cos\theta\sin\phi
- \hat{J}_{z}^3\sin^3\theta\nonumber\\
 &+& (\hat{J}_{x}\hat{J}_{y} + \hat{J}_{y}\hat{J}_{x})\hat{J}_{x}\cos^3\theta\sin\phi\cos^2\phi\nonumber\\
&+& (\hat{J}_{x}\hat{J}_{y} + \hat{J}_{y}\hat{J}_{x})\hat{J}_{y}\cos^3\theta\sin^2\phi\cos\phi\nonumber\\
&-& (\hat{J}_{x}\hat{J}_{y} + \hat{J}_{y}\hat{J}_{x})\hat{J}_{z}\sin\theta\cos^2\theta\sin\phi\cos\phi\nonumber\\
&-& (\hat{J}_{x}\hat{J}_{z} + \hat{J}_{z}\hat{J}_{x})\hat{J}_{x}
\sin\theta\cos^2\theta\cos^2\phi\nonumber\\
&-& (\hat{J}_{x}\hat{J}_{z} + \hat{J}_{z}\hat{J}_{x})\hat{J}_{y}\sin\theta\cos^2\theta
\sin\phi\cos\phi \nonumber\\
&+& (\hat{J}_{x}\hat{J}_{z} + \hat{J}_{z}\hat{J}_{x})\hat{J}_{z}\sin^2\theta
\cos\theta\cos\phi \nonumber\\
&-& (\hat{J}_{y}\hat{J}_{z} + \hat{J}_{z}\hat{J}_{y})\hat{J}_{x}
\sin\theta
\cos^2\theta\sin\phi\cos\phi\nonumber\\
&-& (\hat{J}_{y}\hat{J}_{z} + \hat{J}_{z}\hat{J}_{y})\hat{J}_{y}
\sin\theta\cos^2\theta\sin^2\phi\nonumber\\
 &+& (\hat{J}_{y}\hat{J}_{z} + \hat{J}_{z}\hat{J}_{y})\hat{J}_{z}
\sin^2\theta\cos\theta\sin\phi.
\label{3.2}
\end{eqnarray} 
We, now, express the above expression in terms of the individual atomic operators $\hat{J}_{1_{x,y,z}}$, $\hat{J}_{2_{x,y,z}}$, and $\hat{J}_{3_{x,y,z}}$. The motivation behind this is to find out how the individual atoms are correlated with each other. For that we use the following results.
\begin{eqnarray}
\hat{J}_{n_{x,y,z}}^2 &=& 1/4,~~ \hat{J}_{n_{x,y,z}}^3 = (1/4)\hat{J}_{n_{x,y,z}}\label{3.3a}\\
\hat{J}_{n_{x}}\hat{J}_{n_{y}} &=& \frac{i}{2}\hat{J}_{n_{z}},~~ \hat{J}_{n_{y}}\hat{J}_{n_{z}} = \frac{i}{2}\hat{J}_{n_{x}}\label{3.3b}\\
\hat{J}_{n_{z}}\hat{J}_{n_{x}} &=& \frac{i}{2}\hat{J}_{n_{y}}
\label{3.3c}
\end{eqnarray}
Using Eqs. (\ref{3.3a}), (\ref{3.3b}), (\ref{3.3c}) and the fact that $\hat{J}_{n_{x}}$, $\hat{J}_{n_{y}}$, and $\hat{J}_{n_{z}}$ anticommute with each other, we evaluate all the operator parts of Eq. (\ref{3.2}) separately and the results are as shown below.

% If in two-column mode, this environment will change to single-column
% format so that long equations can be displayed. Use
% sparingly.
\begin{equation}
\hat{J}_{x}^3 = \frac{7}{4}\bigg(\hat{J}_{1_{x}} + \hat{J}_{2_{x}} + \hat{J}_{3_{x}}\bigg) + 6\hat{J_{1_x}}\hat{J_{2_x}}\hat{J_{3_x}}
\label{3.4}
\end{equation}
\begin{eqnarray}
\hat{J}_{x}^2\hat{J}_{y} &=& \frac{3}{4}\bigg{(}\hat{J}_{1_{y}}+ \hat{J}_{2_{y}}+\hat{J}_{3_{y}}\bigg{)} + i\hat{J_{1_z}}\hat{J_{2_x}} + i\hat{J}_{1_{x}}\hat{J}_{2_{z}}\nonumber\\
&+& i\hat{J}_{1_{z}}\hat{J}_{3_{x}} + i\hat{J}_{1_{x}}\hat{J}_{3_{z}} + i\hat{J}_{2_{z}}\hat{J}_{3_{x}} + i\hat{J}_{2_{x}}\hat{J}_{3_{z}} + \nonumber\\
&&2\hat{J}_{1_{x}}\hat{J}_{2_{x}}\hat{J}_{3_{y}} + 2\hat{J}_{1_{x}}\hat{J}_{2_{y}}\hat{J}_{3_{x}} + 2\hat{J}_{1_{y}}\hat{J}_{2_{x}}\hat{J}_{3_{x}}\nonumber\\
\label{3.5}
\end{eqnarray}

\begin{eqnarray}
\hat{J}_{x}^2\hat{J}_{z} &=& \frac{3}{4}\bigg{(}\hat{J}_{1_{z}}+ \hat{J}_{2_{z}}+\hat{J}_{3_{z}}\bigg{)} - i\hat{J_{1_y}}\hat{J_{2_x}} - i\hat{J}_{1_{x}}\hat{J}_{2_{y}}\nonumber\\
&-& i\hat{J}_{1_{y}}\hat{J}_{3_{x}} - i\hat{J}_{1_{x}}\hat{J}_{3_{y}} - i\hat{J}_{2_{y}}\hat{J}_{3_{x}} - i\hat{J}_{2_{x}}\hat{J}_{3_{y}} + \nonumber\\
&&2\hat{J}_{1_{x}}\hat{J}_{2_{x}}\hat{J}_{3_{z}} + 2\hat{J}_{1_{x}}\hat{J}_{2_{z}}\hat{J}_{3_{x}} + 2\hat{J}_{1_{z}}\hat{J}_{2_{x}}\hat{J}_{3_{x}},\nonumber\\
\label{3.6}
\end{eqnarray}

\begin{eqnarray}
\hat{J}_{y}^2\hat{J}_{x} &=& \frac{3}{4}\bigg{(}\hat{J}_{1_{x}}+ \hat{J}_{2_{x}}+\hat{J}_{3_{x}}\bigg{)} - i\hat{J_{1_z}}\hat{J_{2_y}}\nonumber\\ 
&-& i\hat{J}_{1_{y}}\hat{J}_{2_{z}}
- i\hat{J}_{1_{z}}\hat{J}_{3_{y}} - i\hat{J}_{1_{y}}\hat{J}_{3_{z}} \nonumber\\
&-& i\hat{J}_{2_{z}}\hat{J}_{3_{y}} - i\hat{J}_{2_{y}}\hat{J}_{3_{z}} + 2\hat{J}_{1_{y}}\hat{J}_{2_{y}}\hat{J}_{3_{x}}\nonumber\\
&+& 2\hat{J}_{1_{y}}\hat{J}_{2_{x}}\hat{J}_{3_{y}} + 2\hat{J}_{1_{x}}\hat{J}_{2_{y}}\hat{J}_{3_{y}},
\label{3.7}
\end{eqnarray}

\begin{equation}
\hat{J}_{y}^3 = \frac{7}{4}\bigg(\hat{J}_{1_{y}} + \hat{J}_{2_{y}} + \hat{J}_{3_{y}}\bigg) + 6\hat{J_{1_y}}\hat{J_{2_y}}\hat{J_{3_y}}
\label{3.8}
\end{equation}

\begin{eqnarray}
\hat{J}_{y}^2\hat{J}_{z} &=& \frac{3}{4}\bigg{(}\hat{J}_{1_{z}}+ \hat{J}_{2_{z}}+\hat{J}_{3_{z}}\bigg{)} + i\hat{J_{1_x}}\hat{J_{2_y}}\nonumber\\ 
&+& i\hat{J}_{1_{y}}\hat{J}_{2_{x}}
+ i\hat{J}_{1_{x}}\hat{J}_{3_{y}} + i\hat{J}_{1_{y}}\hat{J}_{3_{x}}\nonumber\\
&+& i\hat{J}_{2_{x}}\hat{J}_{3_{y}} + i\hat{J}_{2_{y}}\hat{J}_{3_{x}} + 
2\hat{J}_{1_{y}}\hat{J}_{2_{y}}\hat{J}_{3_{z}}\nonumber\\
&+& 2\hat{J}_{1_{y}}\hat{J}_{2_{z}}\hat{J}_{3_{y}} + 2\hat{J}_{1_{z}}\hat{J}_{2_{y}}\hat{J}_{3_{y}},
\label{3.9}
\end{eqnarray}

\begin{eqnarray}
\hat{J}_{z}^2\hat{J}_{x} &=& \frac{3}{4}\bigg{(}\hat{J}_{1_{x}}+ \hat{J}_{2_{x}}+\hat{J}_{3_{x}}\bigg{)} + i\hat{J_{1_y}}\hat{J_{2_z}}\nonumber\\
&+& i\hat{J}_{1_{y}}\hat{J}_{3_{z}}
+ i\hat{J}_{2_{y}}\hat{J}_{3_{z}} + i\hat{J}_{1_{z}}\hat{J}_{2_{y}}\nonumber\\ 
&+& i\hat{J}_{1_{z}}\hat{J}_{3_{y}}
+ i\hat{J}_{2_{z}}\hat{J}_{3_{y}} + 
2\hat{J}_{1_{z}}\hat{J}_{2_{z}}\hat{J}_{3_{x}}\nonumber\\
&+& 2\hat{J}_{1_{z}}\hat{J}_{2_{x}}\hat{J}_{3_{z}} + 2\hat{J}_{1_{x}}\hat{J}_{2_{z}}\hat{J}_{3_{z}},
\label{3.10}
\end{eqnarray}

\begin{eqnarray}
\hat{J}_{z}^2\hat{J}_{y} &=& \frac{3}{4}\bigg{(}\hat{J}_{1_{y}}+ \hat{J}_{2_{y}}+\hat{J}_{3_{y}}\bigg{)} - i\hat{J_{1_x}}\hat{J_{2_z}}\nonumber\\
&-& i\hat{J}_{1_{x}}\hat{J}_{3_{z}}
- i\hat{J}_{1_{z}}\hat{J}_{2_{x}} - i\hat{J}_{1_{z}}\hat{J}_{3_{x}}\nonumber\\
&-& i\hat{J}_{2_{x}}\hat{J}_{3_{z}} - i\hat{J}_{2_{z}}\hat{J}_{3_{x}} + 
2\hat{J}_{1_{y}}\hat{J}_{2_{z}}\hat{J}_{3_{z}}\nonumber\\
&+& 2\hat{J}_{1_{z}}\hat{J}_{2_{z}}\hat{J}_{3_{y}} + 2\hat{J}_{1_{z}}\hat{J}_{2_{y}}\hat{J}_{3_{z}},
\label{3.11}
\end{eqnarray}

\begin{equation}
\hat{J}_{z}^3 = \frac{7}{4}\bigg(\hat{J}_{1_{z}} + \hat{J}_{2_{z}} + \hat{J}_{3_{z}}\bigg) + 6\hat{J_{1_z}}\hat{J_{2_z}}\hat{J_{3_z}}
\label{3.12}
\end{equation}

\begin{eqnarray}
\hat{J}_{x}\hat{J}_{y}\hat{J}_{x} &=& \frac{1}{4}\bigg(\hat{J}_{1_{y}} + \hat{J}_{2_{y}} + \hat{J}_{3_{y}}\bigg) + 2\hat{J_{1_x}}\hat{J_{2_y}}\hat{J_{3_x}}\nonumber\\
&+& 2\hat{J_{1_x}}\hat{J_{2_x}}\hat{J_{3_y}} + 2\hat{J_{1_y}}\hat{J_{2_x}}\hat{J_{3_x}} 
\label{3.13}
\end{eqnarray}

\begin{eqnarray}
\hat{J}_{y}\hat{J}_{x}^2 &=& \frac{3}{4}\bigg{(}\hat{J}_{1_{y}}+ \hat{J}_{2_{y}}+\hat{J}_{3_{y}}\bigg{)} - i\hat{J_{1_z}}\hat{J_{2_x}}  \nonumber\\
&-&i\hat{J}_{1_{z}}\hat{J}_{3_{x}}
- i\hat{J}_{1_{x}}\hat{J}_{2_{z}} - i\hat{J}_{1_{x}}\hat{J}_{3_{z}}\nonumber\\
&-&  i\hat{J}_{2_{z}}\hat{J}_{3_{x}} - i\hat{J}_{2_{x}}\hat{J}_{3_{z}} + 2\hat{J}_{1_{y}}\hat{J}_{2_{x}}\hat{J}_{3_{x}}\nonumber\\  
&+& 2\hat{J}_{1_{x}}\hat{J}_{2_{y}}\hat{J}_{3_{x}} + 2\hat{J}_{1_{x}}\hat{J}_{2_{x}}\hat{J}_{3_{y}},
\label{3.14}
\end{eqnarray}

\begin{eqnarray}
\hat{J}_{x}\hat{J}_{y}^2 &=& \frac{3}{4}\bigg{(}\hat{J}_{1_{x}}+ \hat{J}_{2_{x}}+\hat{J}_{3_{x}}\bigg{)} + i\hat{J_{1_z}}\hat{J_{2_y}}  \nonumber\\
&+& i\hat{J}_{1_{z}}\hat{J}_{3_{y}}
+ i\hat{J}_{1_{y}}\hat{J}_{2_{z}} + i\hat{J}_{1_{y}}\hat{J}_{3_{z}}\nonumber\\
&+& i\hat{J}_{2_{z}}\hat{J}_{3_{y}} + i\hat{J}_{2_{y}}\hat{J}_{3_{z}} 
+ 2\hat{J}_{1_{x}}\hat{J}_{2_{y}}\hat{J}_{3_{y}}\nonumber\\
&+& 2\hat{J}_{1_{y}}\hat{J}_{2_{x}}\hat{J}_{3_{y}} + 2\hat{J}_{1_{y}}\hat{J}_{2_{y}}\hat{J}_{3_{x}},
\label{3.15}
\end{eqnarray}

\begin{eqnarray}
\hat{J}_{y}\hat{J}_{x}\hat{J}_{y} &=& \frac{1}{4}\bigg(\hat{J}_{1_{x}} + \hat{J}_{2_{x}} + \hat{J}_{3_{x}}\bigg) + 2\hat{J_{1_y}}\hat{J_{2_x}}\hat{J_{3_y}}\nonumber\\
&+& 2\hat{J_{1_y}}\hat{J_{2_y}}\hat{J_{3_x}} + 2\hat{J_{1_x}}\hat{J_{2_y}}\hat{J_{3_y}} 
\label{3.16}
\end{eqnarray}

\begin{eqnarray}
\hat{J}_{x}\hat{J}_{y}\hat{J}_{z} &=& i\frac{3}{8} + i\hat{J_{1_z}}\hat{J_{2_z}} + i\hat{J}_{1_{z}}\hat{J}_{3_{z}}- i\hat{J}_{1_{y}}\hat{J}_{2_{y}}\nonumber\\
&+& i\hat{J}_{1_{x}}\hat{J}_{2_{x}} - i\hat{J}_{1_{y}}\hat{J}_{3_{y}}
+ i\hat{J}_{1_{x}}\hat{J}_{3_{x}}\nonumber\\
&+& i\hat{J}_{2_{z}}\hat{J}_{3_{z}}
- i\hat{J}_{2_{y}}\hat{J}_{3_{y}}
+ i\hat{J}_{2_{x}}\hat{J}_{3_{x}}\nonumber\\
&+& \hat{J}_{1_{x}}\hat{J}_{2_{y}}\hat{J}_{3_{z}} + 
\hat{J}_{1_{x}}\hat{J}_{2_{z}}\hat{J}_{3_{y}}
+ \hat{J}_{1_{y}}\hat{J}_{2_{x}}\hat{J}_{3_{z}}\nonumber\\
&+& \hat{J}_{1_{z}}\hat{J}_{2_{x}}\hat{J}_{3_{y}} + \hat{J}_{1_{y}}\hat{J}_{2_{z}}\hat{J}_{3_{x}}
+ \hat{J}_{1_{z}}\hat{J}_{2_{y}}\hat{J}_{3_{x}},\nonumber\\
\label{3.17}
\end{eqnarray}

\begin{eqnarray}
\hat{J}_{y}\hat{J}_{x}\hat{J}_{z} &=& -i\frac{3}{8} - i\hat{J_{1_z}}\hat{J_{2_z}} - i\hat{J}_{1_{z}}\hat{J}_{3_{z}}
+ i\hat{J}_{1_{x}}\hat{J}_{2_{x}}\nonumber\\
&-& i\hat{J}_{1_{y}}\hat{J}_{2_{y}} - i\hat{J}_{1_{y}}\hat{J}_{3_{y}} + i\hat{J}_{1_{x}}\hat{J}_{3_{x}}  \nonumber\\
&-& i\hat{J}_{2_{z}}\hat{J}_{3_{z}}
- i\hat{J}_{2_{y}}\hat{J}_{3_{y}} + i\hat{J}_{2_{x}}\hat{J}_{3_{x}}\nonumber\\
&+& \hat{J}_{1_{y}}\hat{J}_{2_{x}}\hat{J}_{3_{z}} + 
\hat{J}_{1_{y}}\hat{J}_{2_{z}}\hat{J}_{3_{x}} + \hat{J}_{1_{x}}\hat{J}_{2_{y}}\hat{J}_{3_{z}}\nonumber\\
&+& \hat{J}_{1_{z}}\hat{J}_{2_{y}}\hat{J}_{3_{x}} + \hat{J}_{1_{x}}\hat{J}_{2_{z}}\hat{J}_{3_{y}}
 + \hat{J}_{1_{z}}\hat{J}_{2_{x}}\hat{J}_{3_{y}},\nonumber\\
\label{3.18}
\end{eqnarray}

\begin{eqnarray}
\hat{J}_{x}\hat{J}_{z}\hat{J}_{x} &=& \frac{1}{4}\bigg(\hat{J}_{1_{z}} + \hat{J}_{2_{z}} + \hat{J}_{3_{z}}\bigg) + 2\hat{J_{1_x}}\hat{J_{2_z}}\hat{J_{3_x}}\nonumber\\
&+& 2\hat{J_{1_x}}\hat{J_{2_x}}\hat{J_{3_z}} + 2\hat{J_{1_z}}\hat{J_{2_x}}\hat{J_{3_x}} 
\label{3.19}
\end{eqnarray}

\begin{eqnarray}
\hat{J}_{z}\hat{J}_{x}^2 &=& \frac{3}{4}\bigg{(}\hat{J}_{1_{z}}+ \hat{J}_{2_{z}}+\hat{J}_{3_{z}}\bigg{)} + i\hat{J_{1_y}}\hat{J_{2_x}}  \nonumber\\
&+& i\hat{J}_{1_{y}}\hat{J}_{3_{x}}
+ i\hat{J}_{1_{x}}\hat{J}_{2_{y}} + i\hat{J}_{2_{y}}\hat{J}_{3_{x}}\nonumber\\
&+& i\hat{J}_{1_{x}}\hat{J}_{3_{y}} + i\hat{J}_{2_{x}}\hat{J}_{3_{y}} + 2\hat{J}_{1_{x}}\hat{J}_{2_{z}}\hat{J}_{3_{x}}\nonumber\\
&+& 2\hat{J}_{1_{x}}\hat{J}_{2_{x}}\hat{J}_{3_{z}} + 2\hat{J}_{1_{z}}\hat{J}_{2_{x}}\hat{J}_{3_{x}},
\label{3.20}
\end{eqnarray}

\begin{eqnarray}
\hat{J}_{x}\hat{J}_{z}\hat{J}_{y} &=& -i\frac{3}{8} - i\hat{J_{1_y}}\hat{J_{2_y}} - i\hat{J}_{1_{y}}\hat{J}_{3_{y}}
+ i\hat{J}_{1_{z}}\hat{J}_{2_{z}}\nonumber\\
&-& i\hat{J}_{1_{x}}\hat{J}_{2_{x}} + i\hat{J}_{1_{z}}\hat{J}_{3_{z}} - i\hat{J}_{1_{x}}\hat{J}_{3_{x}}  \nonumber\\
&-& i\hat{J}_{2_{y}}\hat{J}_{3_{y}}
+ i\hat{J}_{2_{z}}\hat{J}_{3_{z}} - i\hat{J}_{2_{x}}\hat{J}_{3_{x}}\nonumber\\
&+& \hat{J}_{1_{x}}\hat{J}_{2_{z}}\hat{J}_{3_{y}} + 
\hat{J}_{1_{x}}\hat{J}_{2_{y}}\hat{J}_{3_{z}} + \hat{J}_{1_{z}}\hat{J}_{2_{x}}\hat{J}_{3_{y}}\nonumber\\
&+& \hat{J}_{1_{y}}\hat{J}_{2_{x}}\hat{J}_{3_{z}} + \hat{J}_{1_{z}}\hat{J}_{2_{y}}\hat{J}_{3_{x}}
 + \hat{J}_{1_{y}}\hat{J}_{2_{z}}\hat{J}_{3_{x}},\nonumber\\
\label{3.21}
\end{eqnarray}

\begin{eqnarray}
\hat{J}_{z}\hat{J}_{x}\hat{J}_{y} &=& i\frac{3}{8} + i\hat{J_{1_y}}\hat{J_{2_y}} + i\hat{J}_{1_{y}}\hat{J}_{3_{y}}
- i\hat{J}_{1_{x}}\hat{J}_{2_{x}}\nonumber\\
&+& i\hat{J}_{1_{z}}\hat{J}_{2_{z}} - i\hat{J}_{1_{x}}\hat{J}_{3_{x}} + i\hat{J}_{1_{z}}\hat{J}_{3_{z}}  \nonumber\\
&+& i\hat{J}_{2_{y}}\hat{J}_{3_{y}}
- i\hat{J}_{2_{x}}\hat{J}_{3_{x}} + i\hat{J}_{2_{z}}\hat{J}_{3_{z}}\nonumber\\
&+& \hat{J}_{1_{z}}\hat{J}_{2_{x}}\hat{J}_{3_{y}}  
+\hat{J}_{1_{z}}\hat{J}_{2_{y}}\hat{J}_{3_{x}} + \hat{J}_{1_{x}}\hat{J}_{2_{z}}\hat{J}_{3_{y}}
\nonumber\\
&+& \hat{J}_{1_{y}}\hat{J}_{2_{z}}\hat{J}_{3_{x}}
+ \hat{J}_{1_{x}}\hat{J}_{2_{y}}\hat{J}_{3_{z}}
 + \hat{J}_{1_{y}}\hat{J}_{2_{x}}\hat{J}_{3_{z}},
\nonumber\\
\label{3.22}
\end{eqnarray}

\begin{eqnarray}
\hat{J}_{x}\hat{J}_{z}^2 &=& \frac{3}{4}\bigg{(}\hat{J}_{1_{x}}+ \hat{J}_{2_{x}}+\hat{J}_{3_{x}}\bigg{)} - i\hat{J_{1_y}}\hat{J_{2_z}} \nonumber\\
&-& i\hat{J}_{1_{y}}\hat{J}_{3_{z}}
- i\hat{J}_{1_{z}}\hat{J}_{2_{y}} - i\hat{J}_{2_{y}}\hat{J}_{3_{z}} - i\hat{J}_{1_{z}}\hat{J}_{3_{y}}\nonumber\\
&-& i\hat{J}_{2_{z}}\hat{J}_{3_{y}}  
 + 2\hat{J}_{1_{x}}\hat{J}_{2_{z}}\hat{J}_{3_{z}} + 2\hat{J}_{1_{z}}\hat{J}_{2_{x}}\hat{J}_{3_{z}}\nonumber\\
 &+& 2\hat{J}_{1_{z}}\hat{J}_{2_{z}}\hat{J}_{3_{x}},
\label{3.23}
\end{eqnarray}

\begin{eqnarray}
\hat{J}_{z}\hat{J}_{x}\hat{J}_{z} &=& \frac{1}{4}\bigg(\hat{J}_{1_{x}} + \hat{J}_{2_{x}} + \hat{J}_{3_{x}}\bigg) + 2\hat{J_{1_z}}\hat{J_{2_x}}\hat{J_{3_z}}\nonumber\\
&+& 2\hat{J_{1_z}}\hat{J_{2_z}}\hat{J_{3_x}} + 2\hat{J_{1_x}}\hat{J_{2_z}}\hat{J_{3_z}} 
\label{3.24}
\end{eqnarray}

\begin{eqnarray}
\hat{J}_{y}\hat{J}_{z}\hat{J}_{x} &=& i\frac{3}{8} + i\hat{J_{1_x}}\hat{J_{2_x}} + i\hat{J}_{1_{x}}\hat{J}_{3_{x}}
- i\hat{J}_{1_{z}}\hat{J}_{2_{z}}\nonumber\\
&+& i\hat{J}_{1_{y}}\hat{J}_{2_{y}}
- i\hat{J}_{1_{z}}\hat{J}_{3_{z}} + i\hat{J}_{1_{y}}\hat{J}_{3_{y}}  \nonumber\\
&+& i\hat{J}_{2_{x}}\hat{J}_{3_{x}}
- i\hat{J}_{2_{z}}\hat{J}_{3_{z}} + i\hat{J}_{2_{y}}\hat{J}_{3_{y}}\nonumber\\ 
&+& 
\hat{J}_{1_{y}}\hat{J}_{2_{z}}\hat{J}_{3_{x}}
+ \hat{J}_{1_{y}}\hat{J}_{2_{x}}\hat{J}_{3_{z}} + \hat{J}_{1_{z}}\hat{J}_{2_{y}}\hat{J}_{3_{x}}\nonumber\\
 &+& \hat{J}_{1_{x}}\hat{J}_{2_{y}}\hat{J}_{3_{z}}
+ \hat{J}_{1_{z}}\hat{J}_{2_{x}}\hat{J}_{3_{y}}
 \nonumber\\
&+& 
\hat{J}_{1_{x}}\hat{J}_{2_{z}}\hat{J}_{3_{y}},
\label{3.25}
\end{eqnarray}

\begin{eqnarray}
\hat{J}_{z}\hat{J}_{y}\hat{J}_{x} &=& -i\frac{3}{8} - i\hat{J_{1_x}}\hat{J_{2_x}} - i\hat{J}_{1_{x}}\hat{J}_{3_{x}}
\nonumber\\
&+& i\hat{J}_{1_{y}}\hat{J}_{2_{y}}
- i\hat{J}_{1_{z}}\hat{J}_{2_{z}} + i\hat{J}_{1_{y}}\hat{J}_{3_{y}}\nonumber\\
 &-& i\hat{J}_{1_{z}}\hat{J}_{3_{z}}
 - i\hat{J}_{2_{x}}\hat{J}_{3_{x}}
+ i\hat{J}_{2_{y}}\hat{J}_{3_{y}} \nonumber\\
 &-& i\hat{J}_{2_{z}}\hat{J}_{3_{z}}
 + \hat{J}_{1_{z}}\hat{J}_{2_{y}}\hat{J}_{3_{x}} 
+ \hat{J}_{1_{z}}\hat{J}_{2_{x}}\hat{J}_{3_{y}}\nonumber\\ 
&+& \hat{J}_{1_{y}}\hat{J}_{2_{z}}\hat{J}_{3_{x}}
 + \hat{J}_{1_{x}}\hat{J}_{2_{z}}\hat{J}_{3_{y}}
\nonumber\\
&+& \hat{J}_{1_{y}}\hat{J}_{2_{x}}\hat{J}_{3_{z}}
+ \hat{J}_{1_{x}}\hat{J}_{2_{y}}\hat{J}_{3_{z}},
\label{3.26}
\end{eqnarray}

\begin{eqnarray}
\hat{J}_{y}\hat{J}_{z}\hat{J}_{y} &=& \frac{1}{4}\bigg(\hat{J}_{1_{z}} + \hat{J}_{2_{z}} + \hat{J}_{3_{z}}\bigg) + 2\hat{J_{1_y}}\hat{J_{2_z}}\hat{J_{3_y}}\nonumber\\
&+& 2\hat{J_{1_y}}\hat{J_{2_y}}\hat{J_{3_z}} + 2\hat{J_{1_z}}\hat{J_{2_y}}\hat{J_{3_y}} 
\label{3.27}
\end{eqnarray}

\begin{eqnarray}
\hat{J}_{z}\hat{J}_{y}^2 &=& \frac{3}{4}\bigg{(}\hat{J}_{1_{z}} + \hat{J}_{2_{z}} + \hat{J}_{3_{z}}\bigg{)} - i\hat{J_{1_x}}\hat{J_{2_y}} \nonumber\\
&-& i\hat{J}_{1_{x}}\hat{J}_{3_{y}}
- i\hat{J}_{1_{y}}\hat{J}_{2_{x}} - i\hat{J}_{2_{x}}\hat{J}_{3_{y}}\nonumber\\
&-& i\hat{J}_{1_{y}}\hat{J}_{3_{x}} - i\hat{J}_{2_{y}}\hat{J}_{3_{x}}  
 + 2\hat{J}_{1_{z}}\hat{J}_{2_{y}}\hat{J}_{3_{y}}\nonumber\\
&+& 2\hat{J}_{1_{y}}\hat{J}_{2_{z}}\hat{J}_{3_{y}} + 
2\hat{J}_{1_{y}}\hat{J}_{2_{y}}\hat{J}_{3_{z}},
\label{3.28}
\end{eqnarray}

\begin{eqnarray}
\hat{J}_{y}\hat{J}_{z}^2 &=& \frac{3}{4}\bigg{(}\hat{J}_{1_{y}}+ \hat{J}_{2_{y}}+\hat{J}_{3_{y}}\bigg{)} + i\hat{J_{1_x}}\hat{J_{2_z}}  \nonumber\\
&+& i\hat{J}_{1_{x}}\hat{J}_{3_{z}}
+ i\hat{J}_{1_{z}}\hat{J}_{2_{x}} + i\hat{J}_{2_{x}}\hat{J}_{3_{z}}\nonumber\\
&+& i\hat{J}_{1_{z}}\hat{J}_{3_{x}} + i\hat{J}_{2_{z}}\hat{J}_{3_{x}} 
 + 2\hat{J}_{1_{y}}\hat{J}_{2_{z}}\hat{J}_{3_{z}}\nonumber\\ &+& 2\hat{J}_{1_{z}}\hat{J}_{2_{y}}\hat{J}_{3_{z}} + 2\hat{J}_{1_{z}}\hat{J}_{2_{z}}\hat{J}_{3_{y}},
\label{3.29}
\end{eqnarray}

\begin{eqnarray}
\hat{J}_{z}\hat{J}_{y}\hat{J}_{z} &=& \frac{1}{4}\bigg(\hat{J}_{1_{y}} + \hat{J}_{2_{y}} + \hat{J}_{3_{y}}\bigg) + 2\hat{J_{1_z}}\hat{J_{2_y}}\hat{J_{3_z}}\nonumber\\
&+& 2\hat{J_{1_z}}\hat{J_{2_z}}\hat{J_{3_y}} + 2\hat{J_{1_y}}\hat{J_{2_z}}\hat{J_{3_z}} 
\label{3.30}
\end{eqnarray}

We see from the above results that, the operators in Eq. (\ref{3.2}) are made up of individual atomic operators, bipartite correlation terms and tripartite correlation terms.
Using Eqs. (\ref{3.4}-\ref{3.30}) in Eq. (\ref{3.2}), we see that all the bipartite correlation terms cancel and we are left with only the individual atomic operators and tripartite correlation terms.
The final expression of Eq. (\ref{3.2}) reduces to

\begin{eqnarray}
\hat{J}_{x^\prime}^3 &=& \frac{7}{4}\bigg(\hat{J}_{1_{x^\prime}}+ \hat{J}_{2_{x^\prime}} + \hat{J}_{3_{x^\prime}} \bigg)\nonumber\\ 
&+& \sum_{\substack{p,q,r={1}\\ {p\ne q\ne r}}}^{3} \bigg( \hat{J}_{p_x}\hat{J}_{q_x}\hat{J}_{r_x}\cos^3\theta\cos^3\phi \nonumber\\
&+& \hat{J}_{p_y}\hat{J}_{q_y}\hat{J}_{r_y}\cos^3\theta\sin^3\phi\nonumber\\
 &-& \hat{J}_{p_z}\hat{J}_{q_z}\hat{J}_{r_z}\sin^3\theta\nonumber\\
&-& 6 \hat{J}_{p_x}\hat{J}_{q_y}\hat{J}_{r_z} \sin\theta\cos^2\theta\sin\phi\cos\phi\nonumber\\
 &+& 3\hat{J}_{p_x}\hat{J}_{q_x}\hat{J}_{r_y}
\cos^3\theta\sin\phi\cos^2\phi\nonumber\\
 &-& 3 \hat{J}_{p_x}\hat{J}_{q_x}\hat{J}_{r_z}\sin\theta\cos^2\theta\cos^2\phi\nonumber\\
&+& 3\hat{J}_{p_x}\hat{J}_{q_y}\hat{J}_{r_y}\cos^3\theta\sin^2\phi\cos\phi\nonumber\\
 &-& 3\hat{J}_{p_y}\hat{J}_{q_y}\hat{J}_{r_z}\sin\theta\cos^2\theta\sin^2\phi\nonumber\\ 
&+& 3 \hat{J}_{p_x}\hat{J}_{q_z}\hat{J}_{r_z}\sin^2\theta\cos\theta\cos\phi
\nonumber\\
 &+& 3 \hat{J}_{p_y}\hat{J}_{q_z}\hat{J}_{r_z}\sin^2\theta\cos\theta\sin\phi \bigg).\nonumber\\
\label{3.30a1}
\end{eqnarray}
The first term on the right hand side is $\frac{7}{4}\hat{J}_{x^\prime}$ whose expectation value over $|\psi_3\rangle$ in this coordinate frame, that is, $(x^\prime, y^\prime, z^\prime)$ is zero. Therefore, we observe that the third order moment $\Delta J_{x^\prime}^3 = \langle \hat{J}_{x^\prime}^3\rangle$ is made up of only the tripartite correlation terms. So, we write
\begin{eqnarray}
\Delta J_{x^\prime}^3 &=& \langle\hat{J}_{x^\prime}^3\rangle = 
\sum_{\substack{p,q,r={1}\\ {p\ne q\ne r}}}^{3} \bigg( \langle\hat{J}_{p_x}\hat{J}_{q_x}\hat{J}_{r_x}
\rangle\cos^3\theta\nonumber\\
&\times&\cos^3\phi
+ \langle\hat{J}_{p_y}\hat{J}_{q_y}\hat{J}_{r_y}\rangle\cos^3\theta\sin^3\phi\nonumber\\ 
&-& \langle\hat{J}_{p_z}\hat{J}_{q_z}\hat{J}_{r_z}\rangle\sin^3\theta
\nonumber\\
&-& 6 \langle\hat{J}_{p_x}\hat{J}_{q_y}\hat{J}_{r_z}\rangle \sin\theta\cos^2\theta\sin\phi\cos\phi\nonumber\\
&+& 3\langle
\hat{J}_{p_x}\hat{J}_{q_x}\hat{J}_{r_y}\rangle\cos^3\theta\sin\phi\cos^2\phi\nonumber\\
 &-& 3\langle \hat{J}_{p_x}\hat{J}_{q_x}\hat{J}_{r_z}\rangle\sin\theta\cos^2\theta\cos^2\phi
\nonumber\\
 &+& 3\langle\hat{J}_{p_x}\hat{J}_{q_y}\hat{J}_{r_y}\rangle\cos^3\theta\sin^2\phi\nonumber\\
&\times&\cos\phi - 3\langle\hat{J}_{p_y}\hat{J}_{q_y}\hat{J}_{r_z}\rangle\sin\theta\cos^2\theta
\nonumber\\
&\times& \sin^2\phi
+ 3 \langle\hat{J}_{p_x}\hat{J}_{q_z}\hat{J}_{r_z}\rangle\sin^2\theta\cos\theta\nonumber\\
&\times& \cos\phi + 3 \langle\hat{J}_{p_y}\hat{J}_{q_z}\hat{J}_{r_z}\rangle
\nonumber\\
&\times&\sin^2\theta\cos\theta\sin\phi \bigg).
\label{3.30a2}
\end{eqnarray}
Now, using the expressions of $\cos\theta$, $\cos\phi$ and $\sin\theta$ and, $\sin\phi$ obtained from Eqs. (\ref{2.5d4}) and (\ref{2.5d5}), Eq. (\ref{3.30a2}) takes the form

\begin{eqnarray}
\Delta J_{x^\prime}^3 &=& 
\frac{1}{|\langle\bf{\hat{J}}\rangle|\big[\langle\hat{J}_x\rangle^2 + \langle\hat{J}_y\rangle^2\big]^{3/2}}\nonumber\\ 
&\times& \sum_{\substack{p,q,r={1}\\ {p\ne q\ne r}}}^{3} \bigg( \langle\hat{J}_{p_x}\hat{J}_{q_x}\hat{J}_{r_x}\rangle\langle\hat{J_z}\rangle^3\langle\hat{J_x}\rangle^3\nonumber\\
&+& \langle\hat{J}_{p_y}\hat{J}_{q_y}\hat{J}_{r_y}\rangle\ \langle\hat{J}_z\rangle^3
\langle\hat{J}_y\rangle^3 - \langle\hat{J}_{p_z}\hat{J}_{q_z}\hat{J}_{r_z}\rangle
\nonumber\\
&\times&\big[\langle\hat{J}_x\rangle^6 + \langle\hat{J}_y\rangle^6 + 3\langle\hat{J}_x\rangle^4\langle\hat{J}_y\rangle^2\nonumber\\  
&+& 
3\langle\hat{J}_x\rangle^2\langle\hat{J}_y\rangle^4  \big]
 - 6 \langle\hat{J}_{p_x}\hat{J}_{q_y}\hat{J}_{r_z}\rangle\nonumber\\
&\times& \big[\langle\hat{J}_x\rangle^2 + \langle\hat{J}_y\rangle^2\big]\langle\hat{J}_z\rangle^2\langle\hat{J}_x\rangle\langle\hat{J}_y\rangle\nonumber\\
 &+& 3\langle
\hat{J}_{p_x}\hat{J}_{q_x}\hat{J}_{r_y}\rangle\langle\hat{J}_z\rangle^3\langle\hat{J}_x\rangle^2\langle\hat{J}_y\rangle\nonumber\\
 &-& 3\langle \hat{J}_{p_x}\hat{J}_{q_x}\hat{J}_{r_z}\rangle
 \big[\langle\hat{J}_x\rangle^2 + \langle\hat{J}_y\rangle^2\big]\langle\hat{J}_x\rangle^2\langle\hat{J}_z\rangle^2\nonumber\\
 &+& 3\langle\hat{J}_{p_x}\hat{J}_{q_y}\hat{J}_{r_y}\rangle\langle\hat{J}_z\rangle^3
\langle\hat{J}_x\rangle\langle\hat{J}_y\rangle^2
\nonumber\\
 &-& 3\langle\hat{J}_{p_y}\hat{J}_{q_y}\hat{J}_{r_z}\rangle\big[\langle\hat{J}_x\rangle^2 + \langle\hat{J}_y\rangle^2\big]\langle\hat{J}_y\rangle^2\langle\hat{J}_z\rangle^2\nonumber\\
 &+& 3 \langle\hat{J}_{p_x}\hat{J}_{q_z}\hat{J}_{r_z}\rangle\langle\hat{J}_x\rangle\langle\hat{J}_z\rangle\big[\langle\hat{J}_x\rangle^4 + \langle\hat{J}_y\rangle^4\nonumber\\ 
&+& 2\langle\hat{J}_x\rangle^2\langle\hat{J}_y\rangle^2\big]
 + 3 \langle\hat{J}_{p_y}\hat{J}_{q_z}\hat{J}_{r_z}\rangle\langle\hat{J}_y\rangle\langle\hat{J}_z\rangle\nonumber\\
&\times& \big[\langle\hat{J}_x\rangle^4 + \langle\hat{J}_y\rangle^4 + 2\langle\hat{J}_x\rangle^2\langle\hat{J}_y\rangle^2\big] \bigg).\nonumber\\
\label{3.30a31}
\end{eqnarray}

Similarly, we can calculate the third order moment of 
$J_{y^\prime}$. Using Eq. (\ref{2.5d2}), (\ref{2.5d4}), (\ref{2.5d5}), (\ref{3.3a}), (\ref{3.3b}), (\ref{3.3c}) and the fact that $\hat{J}_{n_{x}}$, $\hat{J}_{n_{y}}$, and $\hat{J}_{n_{z}}$ anticommute with each other, we obtain the third order moment in 
$J_{y^\prime}$ as
\begin{eqnarray}
\Delta J_{y^\prime}^3 &=& 
\frac{1}{\big[\langle\hat{J}_x\rangle^2 + \langle\hat{J}_y\rangle^2\big]^{3/2}} 
\sum_{\substack{p,q,r={1}\\ {p\ne q\ne r}}}^{3}\nonumber\\
&&\bigg( -\langle\hat{J}_{p_x}\hat{J}_{q_x}\hat{J}_{r_x}\rangle\langle\hat{J_y}\rangle^3 + \langle\hat{J}_{p_y}\hat{J}_{q_y}\hat{J}_{r_y}\rangle \nonumber\\
&\times&\langle\hat{J}_x\rangle^3
+ 3\langle
\hat{J}_{p_x}\hat{J}_{q_x}\hat{J}_{r_y}\rangle\langle\hat{J}_x\rangle\langle\hat{J}_y\rangle^2 \nonumber\\
&-& 3\langle\hat{J}_{p_x}\hat{J}_{q_y}\hat{J}_{r_y}\rangle\langle\hat{J}_x\rangle^2\langle\hat{J}_y\rangle\bigg) 
\label{3.31}
\end{eqnarray}
While calculating $\Delta J_{y^\prime}^3$, we find that all the bipartite correlation terms cancel out and it is solely made up of tripartite correlation terms like the previous case.

\subsection{8 ACKNOWLEDGMENTS}
I am grateful to Sunandan Gangopadhyay for sending me the paper in Ref. \cite{Hong}.

\subsection {9 FUNDING AND/OR CONFLICTS OF INTERESTS/COMPETING INTERESTS}
The author declares no potential conflict of interests/competing interests.

This work has been done without any kind of financial support or fund/grant from any kind of organization or agency or any individual.

\end{document}